\documentclass[showpacs,oneside,twocolumn,prl,amsmath,amssymb,superscriptaddress]{revtex4-1}

\usepackage{cases}
\usepackage{amsmath}
\usepackage{amssymb}
\usepackage{amsfonts}
\usepackage{amssymb}
\usepackage{dcolumn}
\usepackage{bm}
\usepackage{bbm}
\usepackage{graphicx}
\usepackage{xcolor}
\usepackage{array}
\usepackage{subfigure}
\usepackage{hyperref}

\hypersetup{colorlinks=true,
            breaklinks=true,
            pdfstartview=Fit,
            linkcolor=blue,
            citecolor=blue,
            urlcolor=blue}

\begin{document}

\title{High-fidelity adiabatic quantum computation using the intrinsic Hamiltonian of a spin system:
Application to the experimental factorization of 291311}

\author{Zhaokai Li}
\affiliation{CAS Key Laboratory of Microscale Magnetic Resonance and Department of Modern Physics, University of Science and Technology of China (USTC), Hefei 230026, China}
\affiliation{Synergetic Innovation Center of Quantum Information and Quantum Physics, USTC, Hefei, China}

\author{Nikesh S. Dattani}
\affiliation{Oxford University, Hertford College, Oxford, OX1 3BW, UK}
\affiliation{Fukui Institute for Fundamental Chemistry, Kyoto University, Kyoto, 606-8103, Japan}

\author{Xi Chen}
\author{Xiaomei Liu}
\author{Hengyan Wang}
\affiliation{CAS Key Laboratory of Microscale Magnetic Resonance and Department of Modern Physics, University of Science and Technology of China (USTC), Hefei 230026, China}

\author{Richard Tanburn}
\affiliation{Oxford University, Hertford College, Oxford, OX1 3BW, UK}

\author{Hongwei Chen}
\affiliation{High Magnetic Field Laboratory, Chinese Academy of Sciences, Hefei 230031, China}

\author{Xinhua Peng} \altaffiliation{xhpeng@ustc.edu.cn}
\affiliation{CAS Key Laboratory of Microscale Magnetic Resonance and Department of Modern Physics, University of Science and Technology of China (USTC), Hefei 230026, China}
\affiliation{Synergetic Innovation Center of Quantum Information and Quantum Physics, USTC, Hefei, China}
\affiliation{Hefei National Laboratory for Physical Sciences at the Microscale, USTC, Hefei, China}

\author{Jiangfeng Du} \altaffiliation{djf@ustc.edu.cn}
\affiliation{CAS Key Laboratory of Microscale Magnetic Resonance and Department of Modern Physics, University of Science and Technology of China (USTC), Hefei 230026, China}
\affiliation{Synergetic Innovation Center of Quantum Information and Quantum Physics, USTC, Hefei, China}
\affiliation{Hefei National Laboratory for Physical Sciences at the Microscale, USTC, Hefei, China}

\begin{abstract}
In previous implementations of adiabatic quantum algorithms using spin systems, the average Hamiltonian method with Trotter's formula was conventionally adopted to generate an effective instantaneous Hamiltonian that simulates an adiabatic passage. However, this approach had issues with the precision of the effective Hamiltonian and with the adiabaticity of the evolution. In order to address these, we here propose and experimentally demonstrate a novel scheme for adiabatic quantum computation by using the intrinsic Hamiltonian of a realistic spin system to represent the problem Hamiltonian while adiabatically driving the system by an extrinsic Hamiltonian directly induced by electromagnetic pulses. In comparison to the conventional method, we observed two advantages of our approach: improved ease of implementation and higher fidelity. As a showcase example of our approach, we experimentally factor 291311, which is larger than any other quantum factorization known.

\end{abstract}

\maketitle

Adiabatic quantum computing (AQC) has been proven to be capable of simulating any circuit-based quantum computer with at most polynomial overhead \cite{AQC_equivalent,AQC_equivalent2}, and therefore has the potential to solve problems that would be very hard on a classical computer. Even when the qubit operations in an adiabatic passage are limited to only involve $\sigma_z$ operators, 2-qubit couplings, limited connectivity, and limited coupling strengths, heuristics demonstrated that discrete optimization problems can be solved using quantum annealing up to 10$^7$ times faster than the fastest single-core classical algorithm implemented \cite{Denchev2015}. In the most recent 2016 study, even the best \textit{parallel} classical algorithms took longer time than quantum annealing to find the ground state for discrete optimization problems of the same kind, with up to 625 binary variables \cite{2016Mandra}.

Therefore, it does not seem fruitless to continue advancing AQC technology to support more general Hamiltonians. Two of the major challenges in present-day quantum annealers using superconducting qubits have been the ability to implement \textit{non-stoquastic} terms, and to implement \textit{$k$-local} terms with $k>2$ in the problem Hamiltonian. Contrarily, annealing in spin systems using Nuclear Magnetic Resonance (NMR), has been successful with \textit{2-local non-stoquastic} terms of types $XX$ and $YY$ (in addition to $ZZ$) \cite{XXandYY}, with \textit{3-local stoquastic} terms \cite{2-3body,Factor_143}, and with \textit{4-local, non-stoquastic} terms of various types \cite{Luo_1,Luo_2}. Since NMR-based annealing has been able to implement $k$-local terms with $k>2$ \textit{without} requiring extra qubits or perturbative gadgets \cite{2-3body, Luo_1, Luo_2}, 5893 qubits would be enough to factor RSA-230 using existing techniques \cite{supp}, while devices such as D-Wave's superconducting flux qubit annealers would require at least 148$\,$776 qubits to quadratize the Hamiltonian for factoring RSA-230 into 2-local form \cite{supp}, and about 5.5 billion physical qubits in total to embed the 148$\,$776 computational qubits onto the current connectivity limitations of, for example, the D-Wave chimera  \cite{Choi2010}. RSA-230 has not yet been factored by a classical computer, and the factorization of RSA-220 took about three years, between 2013 and 2016 \cite{rsa220}.

However, all implementations to date of AQC algorithms using spin systems (of which we are aware), adopted the average Hamiltonian method \cite{average_H} where the adiabatic evolution of the Hamiltonian is decomposed into a series of quantum gates or control pulses. Although the quantum circuits used there have the same propagators as required, the system of qubits exits the ground state during certain segments of the quantum circuit. As the number of qubits grows very large, so does the number of available low-lying excited states into which the system can spuriously get trapped. Furthermore, the amount of control pulses required in the experiments also grows rapidly with the number of steps in an adiabatic passage, leading to an experimental implementation that is vulnerable against noise. These disadvantages have posed a threat to the scalability of spin-based AQC for problems involving thousands of qubits.

In this Letter we demonstrate how to overcome this problem by making use of the intrinsic Hamiltonian of the physical system. The initial Hamiltonian where the adiabatic passage starts is induced by the Hamiltonian of radio frequency pulses, while the problem Hamiltonian where the passage ends is approximated by the intrinsic Hamiltonian of a nuclear spin system. By driving the quantum system through the adiabatic passage faithfully, we do not allow it to escape from the ground state at any moment during the entire process. In this way, the amount of control pulses required is reduced greatly and the experimental implementation is also more robust against noise. The implementation of this technique was made possible due to increased flexibility in the problem Hamiltonian due to Energy Landscape Manipulation (ELM) \cite{{ELM2015}}, which is a fully scalable way to transform a Hamiltonian of computational interest, into another Hamiltonian with the \textit{same unique ground state(s)}, but much more amenable to experimental implementation.

As an example, we report an experimental prime factorization of $N=291311$. Although the well-known Shor algorithm \cite{shors'} has already been demonstrated in different physical systems \cite{shor_NMR,shor_optics,shor_optics2,shor_photonic_chip,shor_phase_qubit,shor_photonic_qubit_recycling,shor_2016_recycling}, the largest number factored by Shor's algorithm is still rather small. Another approach of quantum factoring is to transform it into a binary optimization problem \cite{Burges2012} and then solve it with quantum annealing \cite{Schutzhold,Factor_21,Schutzhold2010,Factor_143}. In our experiment, the prime factors of 291311 are measured to be 523 and 557 at the end of the adiabatic evolution.

First, we describe the general framework for prime factorization as follows. Suppose that the integer $N$ is the number that needs to be factored, while $p$ and $q$ are the prime factors, {\it i.e.}, $N = p \times q$. Here, the factors $p$ and $q$ can be denoted in binary form as $\{1p_mp_{m-1}...p_2p_11\}_\textrm{bin}$ for $p=2^{m+1}+\sum_{i = 1}^{m}p_i \times 2^i+1$ and $\{1q_nq_{n-1}...q_2q_11\}_\textrm{bin}$ for $q$. In this form, the factorization problem is to find the values of $p_1,...,p_m,q_1,...,q_n$ that meet the restriction $N = p \times q$. Recent work has shown that the $m+n$ variables can be reduced to a significantly smaller number of variables \cite{arxiv_291311}. For example, the factorization problem of $N=291311$ reduces to the equations \cite{arxiv_291311}:
\begin{equation}\label{pqs}
\begin{split}
p_1+q_1=1\\
p_2+q_2=1\\
p_5+q_5=1\\
p_1q_2+p_2q_1=1\\
p_2q_5+p_5q_2=0\\
p_5q_1+p_1q_5=1,
\end{split}
\end{equation}
where the binary form of the factors are $p=\{1000p_501p_2p_11\}_\textrm{bin}$ and $q=\{1000q_501q_2q_11\}_\textrm{bin}$. Since the first three equations imply that $p_i=1-q_i$ for $i=1,2,5$, the equations become:
\begin{equation}\label{qs_bin1}
\begin{split}
q_1+q_2-2q_1q_2=1\\
q_2+q_5-2q_2q_5=0\\
q_1+q_5-2q_1q_5=1,
\end{split}
\end{equation}
which form a 3-variable binary optimization problem. The values of ${q_1,q_2,q_5}$ satisfying Eq.\ (\ref{qs_bin1}), which represent the solution to the factorization problem of $N=291311$, are encoded in the ground-state of the AQC problem Hamiltonian:
\begin{equation}
\label{H_hatBeforeELM}
\begin{split}
H_p = &(\hat q_1+\hat q_2-2\hat q_1\hat q_2-1)^2+ (\hat q_2+\hat q_5-2\hat q_2\hat q_5)^2 \\
&+ (\hat q_1+\hat q_5-2\hat q_1\hat q_5-1)^2.
\end{split}
\end{equation}

\noindent Here, the variables $q_1,q_2 $ and $ q_5$ were mapped into qubit operators that can be written as $\hat q_1=\frac{1-\sigma_z^1}{2}$, $\hat q_2=\frac{1-\sigma_z^2}{2}$, $\hat q_5=\frac{1-\sigma_z^3}{2}$ where $\sigma_{x,y,z}^i$ denotes a Pauli operator acting on the $i^{\rm{th}}$ qubit. 

It can be rather hard to construct a system of spins that have precisely the energies and couplings demanded by the problem Hamiltonian in AQC (e.g. Eq.\ (\ref{H_hatBeforeELM})). So in previous AQC work, the average Hamiltonian method was adopted to use a series of control pulses to mimic such Hamiltonians. Even under ideal conditions (e.g. no noise, and no decoherence), these control pulses can allow the system of qubits to exit the ground state, at which point more control pulses are used to return the system back to the ground state. In this work we avoid this deviation from pure adiabaticity by transforming the Hamiltonian of Eq.\ (\ref{H_hatBeforeELM}) into one which has the \textit{same ground state}, but whose  energies and coupling strengths correspond very closely to a system which is physically realizable and easy enough to control adiabatically. We start by noticing that if we introduce \textit{positive}-valued parameters $\alpha,\beta$ and $\gamma$, the Hamiltonian of Eq.\ (\ref{H_hatBeforeELM}) has the same ground state as the more flexible Hamiltonian below:

\begin{equation}\label{H_hat}
\begin{split}
H_p = &\alpha (\hat q_1+ \hat q_2-2 \hat q_1 \hat q_2-1)^2+ \beta ( \hat q_2+ \hat q_5-2 \hat q_2 \hat q_5)^2 \\ &+ \gamma ( \hat q_1+ \hat q_5-2 \hat q_1 \hat q_5-1)^2.
\end{split}
\end{equation}

This is a specific Hamiltonian transformation within a much more general scheme of energy landscape manipulation (ELM) techniques introduced in \cite{ELM2015}, which allow a Hamiltonian to be transformed into a new one that has the same ground state, but different gaps between the ground state and first excited state, different numbers of low-lying excited states, different coupling strengths, etc.

Using the Pauli operator representation described above, we re-write the Hamiltonian again without changing the ground state:
\begin{equation}\label{H_p}
H_p=\frac{1}{2} \left( \alpha \sigma_z^1 \sigma_z^2 - \beta \sigma_z^2 \sigma_z^3 + \gamma \sigma_z^1 \sigma_z^3 \right),
\end{equation}
where we have neglected a constant term since it will not affect the form of the ground-state, but kept the factor of 1/2 for reasons that will become apparent when we describe the experiments. 

\begin{figure}[hbtp]
 \centering
 \includegraphics[width= 1 \columnwidth]{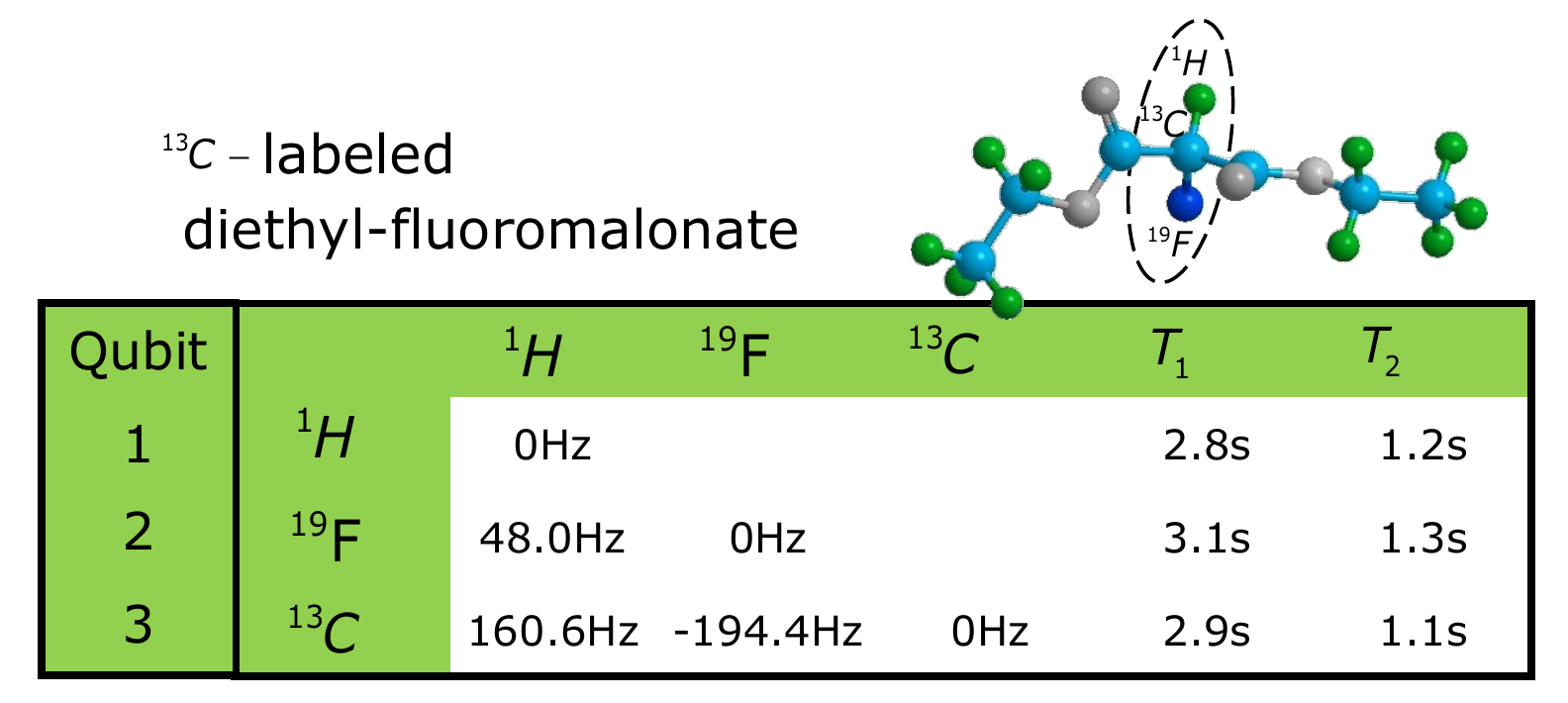}
 \caption{The three qubit NMR quantum processor consists of $^{13}$C-labeled diethyl-fluoromalonate dissolved in d-chloroform. The scalar couplings $J_{jk}$ between nuclear spins are below the diagonal in the table. The Larmor frequencies, which are on the diagonal, reduce to zero in the on-resonance rotating frame.}
 \label{sample}
\end{figure}

We choose the initial Hamiltonian of the adiabatic process to be:
\begin{equation}
{H_0} = {\sigma_x^1} + {\sigma_x^2} + {\sigma_x^3},
\end{equation}
with ground-state $|\phi \rangle _0={(\frac{1}{{\sqrt 2 }}(|0 \rangle  - |1 \rangle ))^{ \otimes 3}}$. The quantum system is first prepared into $|\phi \rangle _0$, and then we evolve it under a time-dependent Hamiltonian $H(s)$ which varies from $H_0$ to $H_p$: $H(s)=(1-s)H_0+sH_p$. If $s$ varies from 0 to 1 slowly enough, the adiabatic theorem suggests that the system will stay in the instantaneous ground state of $H(s)$. At the end of the adiabatic evolution, the system will be in the ground state of $H_p$ which encodes the solution of the factorization problem. In principle we can now construct a physical system with energies and adjustable coupling strengths corresponding to our problem Hamiltonian (as is done in the superconducting chimeras of D-Wave, for example), but our ELM coefficients make $H_p$ so flexible that we can actually realize it with a naturally occurring quantum mechanical system. In particular, the nuclear spins of the atoms in diethyl-fluoromalonate which we have highlighted in Fig.\ \ref{sample}, have coupling strengths of roughly 1.2, -4.9, and 4 relative to 40 Hz. We therefore choose the ELM coefficients as $\alpha=1.2, \beta=4.9, \gamma=4$, which makes the problem Hamiltonian easy to simulate physically with a stable molecule (although in general if we did not have this luxury we could also construct such a quantum mechanical system arbitrarily by using adjustable couplers as done in superconducting systems).

\begin{figure}[hbtp]
 \centering
 \includegraphics[width=1 \columnwidth]{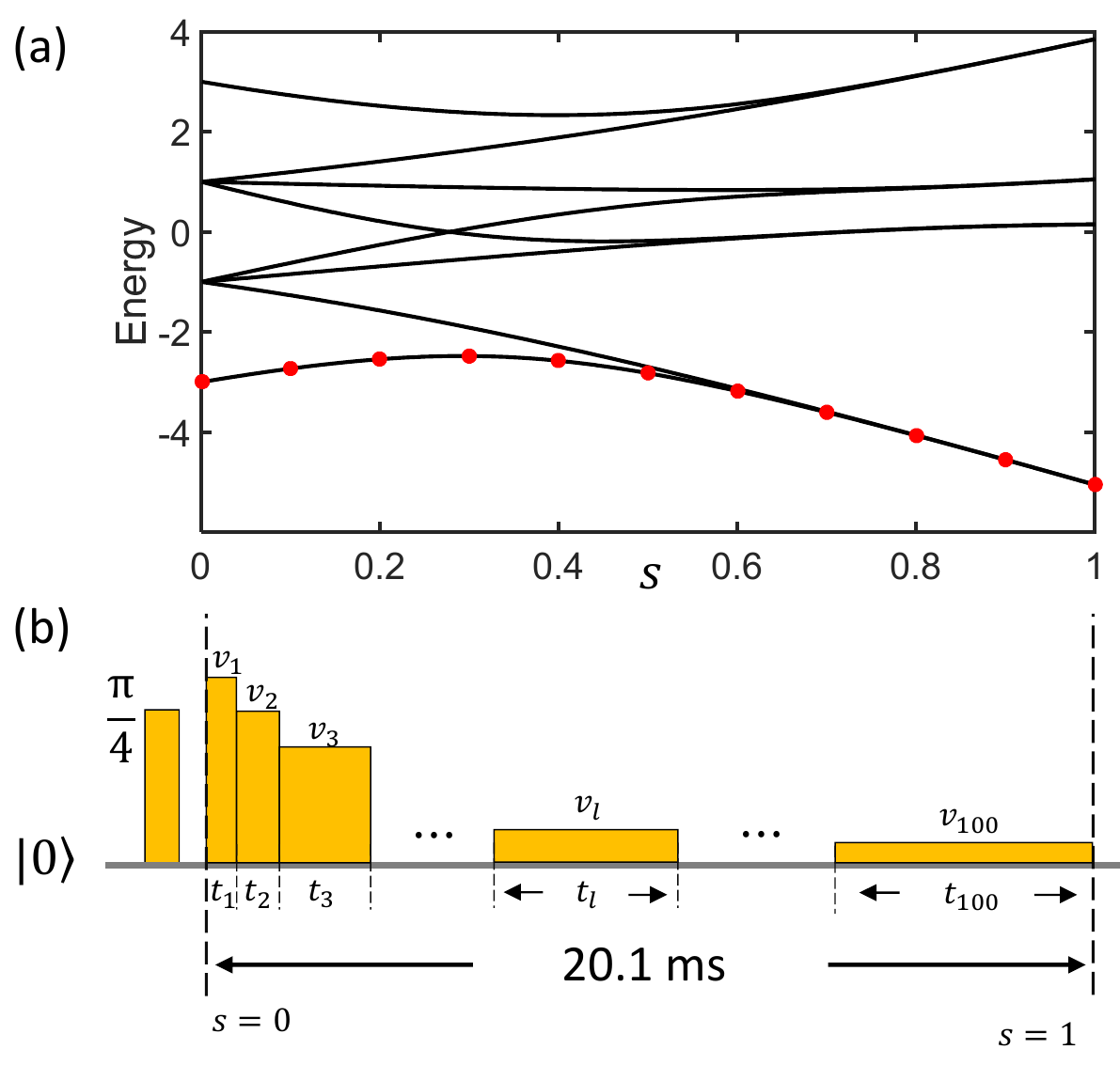}
 \caption{The adiabatic factorization of $N=291311$. (a) The energy levels of the Hamiltonian $H(s)$ in Eq.\ (\ref{Hs}), when $s$ varies from $0$ to $1$. The red dots represent 11 different stages of the adiabatic evolution where we measured the populations in the computational basis. (b) The radio-frequency pulse sequence for the adiabatic factorization. Here the same pulse sequence is applied to three qubits simultaneously. The first pulse indicated with ${\pi  \over 4}$ represents the operation $e^{(i\sigma_y\pi/4)}$. The following shaped pulse represents the adiabatic evolution from $H_0$ to problem Hamiltonian $H_p$. It consists of 100 slices of pulses with different durations ($t_l$) and amplitudes ($\nu_l$) shown in Eqs.\ (\ref{slice_amp}-\ref{slice_t}). Note that these slices are not drawn to scale.}
 \label{circuit}
\end{figure}

The instantaneous Hamiltonian $H(s)$ is now given by:

\begin{equation}\label{Hs}
\footnotesize{
H(s) = (1 - s)\underbrace {(\sigma _x^1 + \sigma _x^2 + \sigma _x^3)}_{{H_0}} + s \cdot \underbrace {{1 \over 2}\left( {1.2\sigma _z^1\sigma _z^2 - 4.9\sigma _z^2\sigma _z^3 + 4\sigma _z^1\sigma _z^3} \right)}_{{H_p}},
}
\end{equation}
and Fig.\ \ref{circuit} shows the energy levels of $H(s)$ when $s$ varies from $0$ to $1$.

We now turn to the experimental process. The experiments were carried on a Bruker AV-400 spectrometer at 304K. The three-qubit quantum processor consists of ${}^{1}H$, ${}^{19}F$, and ${}^{13}C$ nuclear spins in the ${}^{13}C$-labeled diethyl-fluoromalonate molecule dissolved in d-chloroform. The intrinsic Hamiltonian of this three-qubit system in the triple-resonance rotating frame is
\begin{equation}
{H_{\rm{intrinsic}}} = \sum\limits_{1 \le j < k \le 3} {\frac{\pi }{2} {J_{jk}}}\sigma_z^j\sigma_z^k,
\end{equation}
where $J_{jk}$ represents the scalar coupling between the $j^{\rm{th}}$ and $k^{\rm{th}}$ spin. The parameters and molecular structure are shown in Fig.\ \ref{sample}, where the three nuclei used as qubits are marked. If the on-resonance radio-frequency fields $B_{\rm rf}^j$  are applied on the $j^{\rm{th}}$ spins along the $x$-axis, the physical system will evolve under the Hamiltonian
\begin{equation}\label{H_NMR}
H_{\rm{phys}} = \pi \sum\limits_{j = 1}^3 {{\gamma ^j}} B_{\rm rf}^j\sigma _x^j + 20\pi (1.2\sigma _z^1\sigma _z^2 - 4.9\sigma _z^2\sigma _z^3 + 4\sigma _z^1\sigma _z^3),
\end{equation}
where $\gamma ^j$ is the gyromagnetic ratio of each nuclear spin. In the experiment, $B_{\rm rf}^j$ is chosen to assure that ${\gamma ^j} B_{\rm rf}^j=\nu$ for $j=1,2,3$, where $\nu$ is the amplitude (in Hz) of the radio-frequency field applied on the three spins simultaneously. If $\nu$ is chosen appropriately, this Hamiltonian will be a good approximation of Eq.\ (\ref{Hs}), up to a constant factor.

The experimental procedure consists of three steps: (1) preparation of the ground state of $H_0$, (2) adiabatic evolution driven by the time-dependent Hamiltonian $H(s)$, (3) measurement of the final state in the computational basis.

Starting from the thermal equilibrium of the NMR system, the line-selective method \cite{pps} is applied to prepare the pseudo-pure state $\rho_{p}$ as:
\begin{equation}
{\rho_{p}} = \frac{{1 - \varepsilon }}{8}I_8 + \varepsilon |000 \rangle  \langle 000|.
\end{equation}
Here $I_8$ represents the $8 \times 8$ identity operator and $\varepsilon \approx {10^{ - 5}} $ the polarization. Then the ground state of $H_0$, {\it i.e.}, ${(\frac{1}{{\sqrt 2 }}(|0 \rangle  - |1 \rangle ))^{ \otimes 3}}$ , is obtained by applying the rotation $e^{(i\sigma_y\pi/4)}$  on three spins simultaneously. Note that the quantum state of the NMR system is in fact a pseudo ground state, owing to the existence of the maximally mixed state $I_8$. Nonetheless, since $I_8$ is unaffected by any unitary transformation, the pseudo ground state will also be driven along the adiabatic passage, and therefore behaves exactly the same as the true ground state.

\begin{figure}[hbtp]
 \centering
 \includegraphics[width= 1 \columnwidth]{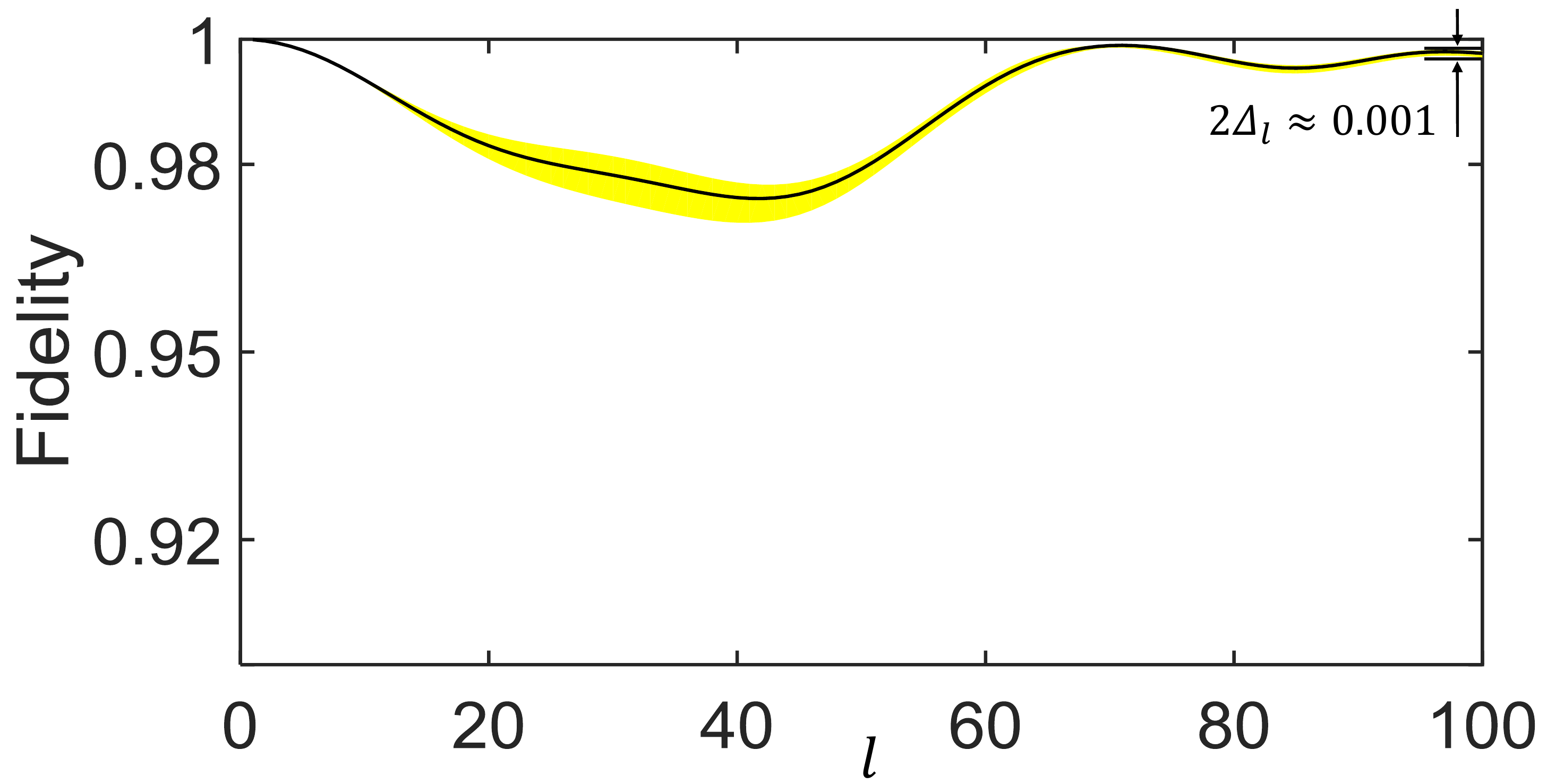}
  \caption{The theoretical fidelities under ideal conditions with the ground state of the instantaneous Hamiltonian during the adiabatic passage (black line). The yellow region labels the standard deviation of fidelities ($2\Delta_l$) around the mean value $F_l$ for each $l$ when all the pulses have random amplitude fluctuations \cite{supp}. The yellow band is asymmetric around the black curve because $F_l$ is slightly different from the black curve.}
  \label{robustness}
\end{figure}

The adiabatic evolution of the system is approximated by $L$ discrete steps with $\tau$ the duration of each step. To ensure the system always stays in the ground state of the instantaneous Hamiltonian, the variation of $H(s)$ should be sufficiently slow, {\it i.e.}, $L \rightarrow \infty$ and $\tau \rightarrow 0$. In the experiment, we choose $\tau=0.05$ and use a linear interpolation with $L=100$, {\it i.e.},  $s_l=0.01 \times l, (l=1,2,...100)$. The numerical simulation shows that the quantum system indeed remains in the ground state with fidelity greater than 0.975 in the entire process (see Fig.\ \ref{robustness}).

For any given $s_l$, the Hamiltonian $H(s)$ can be approximated experimentally by Eq.\ (\ref{H_NMR}) with the amplitude of the radio frequency pulse (in Hz) as:
\begin{equation}\label{slice_amp}
{\nu_l} = 40 \left(\frac{1 - s_l}{s_l}\right).
\end{equation}
Furthermore, for each $l$, the evolution of $H(s)$ over the  duration $\tau$ is simulated by applying the pulse for a period of time (measured in seconds):
\begin{equation}\label{slice_t}
{t_l} = \frac{\tau \cdot s_l }{{40 \pi }}.
\end{equation}
In the experiment, a series of $100$ slices of radio frequency pulses are applied to simulate the $100$-step quantum adiabatic evolution. The entire experimental time of the adiabatic evolution is about 20.1$\,$ms, with pulse sequences shown in Fig.\ \ref{circuit}.

\begin{figure}[hbtp]
 \centering
 \includegraphics[width=0.8 \columnwidth]{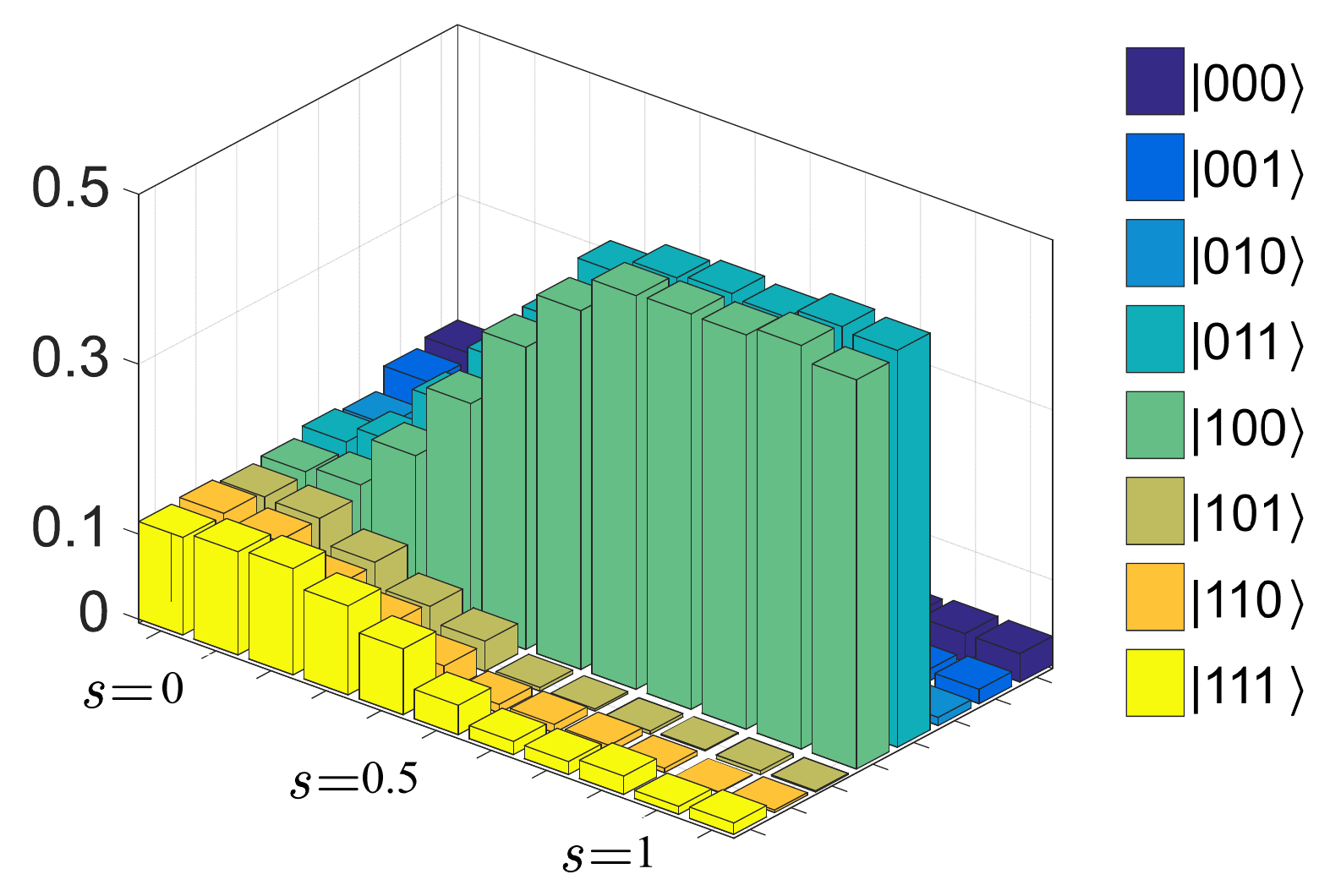}
 \caption{The variation of the populations in the computational basis of our experimental system during the adiabatic evolution with $s$ varying from 0 to 1. The initial state has an equal distribution in the computational basis when $s=0$. At the end of the adiabatic process where $s=1$, the system finally stays on the superposition of $|011\rangle$ and $|100\rangle$, which indicates that the solution of the 3-variable equations (Eq.\ (\ref{qs_bin1})) is $\{q_1=0,q_2=1,q_5=1\}$ or $\{q_1=1, q_2=0, q_5=0\}$.}
 \label{expprocess}
\end{figure}

Finally, the system will stay in the ground state of the problem Hamiltonian $H_p$ with a high fidelity. Projective measurements in the computational basis can be done to give us information about the system's state. In the experiment, three readout pulses are applied to reconstruct the diagonal elements of the density matrix of the final state, {\it i.e.}, the populations in the computational basis. The experimentally measured population of the final state is shown in Fig.\ \ref{expprocess}, labeled with $s=1$. This result shows that the ground state is mainly in the superposition of $|011\rangle$ and $|100\rangle$, indicating the solution of the 3-variable equations (Eq.\ (\ref{qs_bin1})) is $\{q_1=0,q_2=1,q_5=1\}$ or $\{q_1=1, q_2=0, q_5=0\}$. As a result, the answer to the factorization problem of $N = 291311$ is $q=\{1000001011\}_\textrm{bin}=523$ or $q=\{1000101101\}_\textrm{bin}=557$. That is to say: $291311=523 \times 557$, which can be verified easily.

To demonstrate the process of adiabatic evolution more clearly, we measured the populations in the computational basis at 11 different stages of the adiabatic evolution, which we labeled as red dots in Fig.\ \ref{circuit}. The change of the population of the quantum system is shown in Fig. \ref{expprocess}.

The tomographically \cite{tomography} reconstructed density matrix of the experimental final state is shown in the Supplemental Material \cite{supp}, with a fidelity of over 0.99 compared to the theoretical prediction, indicating a high accuracy of the quantum adiabatic evolution in our experiment. The errors mainly come from the imperfections of the pseudo-pure state and decoherence effects.

The inaccuracies of control pulses are mostly eliminated because of the robustness of the adiabatic evolution in our experiment. If all the radio-frequency pulses $(\nu_l)$ have random fluctuations which have a Gaussian distribution with expected value 0 and standard deviation $0.05\nu_l$, the adiabatic passage will also have a random variation. We numerically analyzed the mean values ($F_{l}$) and standard deviations ($\Delta_l$) of the fidelities after each step of the adiabatic passage. The yellow band in Fig.\ \ref{robustness} represents the region $ [F_{l}-\Delta_{l}, F_{l}+\Delta_{l}]$ for different $l$. Although all the pulses have fluctuations of around $5\%$, the standard deviation of the fidelity of the final state is less than $0.001$. As a comparison, if the average Hamiltonian method is used, the standard deviation is around $0.015$ under the influence of the same noise (see Supplemental Material for details \cite{supp}). Furthermore, in the average Hamiltonian method, the quantum system exits the ground state during certain segments (see Fig. S2 in the Supplemental Material \cite{supp}, particularly the periods during which the fidelity with the ideal ground state falls below 0.25).

Therefore, in this Letter we have demonstrated an experimental method for quantum adiabatic computing in spin systems that does not use control pulses that drive the system out of the ground state. In the experiment, the intrinsic Hamiltonian of a realistic quantum system is used to approximate the problem Hamiltonian while an extrinsic Hamiltonian is induced to drive the quantum system to evolve along the adiabatic passage. Compared with the traditional average Hamiltonian method, the desired ground state in our approach is obtained with a much greater fidelity and the experimental realization of the adiabatic evolution is also more robust against noise. The methods we have used in this experiment can be applied to other spin-based AQC architectures which may have various advantages in terms of scaling to larger numbers of qubits, such as NV-centers \cite{NVcenter} for example.

This work is supported by National Key Basic Research Program of China (2013CB921800 and 2014CB848700), the National Natural Science Foundation of China (Grants No.\ 11425523, No.\ 11375167, No.\ 11575173, No.\ 11374308, No.\ U1632157 and No.\ 11227901), the Strategic Priority Research Program (B) of the CAS (Grant No.\ XDB01030400) and Key Research Program of Frontier Sciences of the CAS (Grant No.\ QYZDY-SSW-SLH004). The authors also thank Zhijin Ke and Nathan Bryans for useful discussions, and Dmitri Iouchtchenko for careful proofreading of the manuscript.

\end{document}